# Micro and nano patternable magnetic carbon

Swati Sharma,[1,2*] Arpad M. Rostas,[3] Neil MacKinnon,[1,2] Lorenzo Bordonali,[1,2] Stefan Weber,[3] Jan G. Korvink[1,2]

[1]Department of Microsystems Engineering-IMTEK, University of Freiburg, Germany

[2]Institute of Microstructure Technology, Karlsruhe Institute of Technology, Germany

[3]Institute of Physical Chemistry, University of Freiburg, Germany

[*]Corresponding author: Institute of Microstructure Technology, Karlsruhe Institute of Technology, Hermann-von-Helmholtz-Platz 1, 76334 Eggenstein-Leopoldshafen, Germany

Email: swati.sharma@kit.edu, Phone: +49 721 60829304

*Carbon is conventionally not associated with magnetism, and much of the discussion of its nanotechnology perspectives appears to be centered on its electron transport properties. Among the few existing examples of magnetic carbon production, none has found a direct applicability in scalable micro and nano fabrication*[1-6] *Here we introduce a paramagnetic form of carbon whose precursor polymers can be lithographically patterned into micro and nano structures prior to pyrolysis. This unreactive and thermally robust material features strong room-temperature paramagnetism owing to a large number of unpaired electrons with restricted mobility, which is achieved by controlling the progression of bond dissociation and formation during pyrolysis. The manufacture of this magnetic carbon, having $(3.97 \pm 0.8) \times 10^{17}$ spins/mg, can immediately benefit a number of spintronic and magnetic MEMS applications*[7,8] *and also shed light on the controversial theories concerning the existence and mechanisms of magnetism in carbon.*[9-11]

Magnetic properties of non-$sp^3$ carbon-based systems have prompted intriguing discussions on the possibility of magnetism in carbon in the recent past.[1-3,11] There is an emerging consensus on the existence of magnetism in $sp$ and $sp^2$ carbon materials, which is supported largely by theoretical studies[12-14] and a few experimental observations.[5,6,15] Existing techniques for manufacturing magnetic carbon rely on expensive and cumbersome fabrication pathways, experience significant processing errors, and are mostly limited to laboratory research.[15-17]

In the case of graphitic carbons, the presence of π-electrons may lead to orientation-dependent weak magnetism.[11,12] Some recently developed methods that aim at obtaining a ubiquitous, direction-independent presence of unpaired electron spins in $sp^2$ carbons rely on inducing radical formation by harsh treatments such as ion irradiation[5] or laser ablation[16] to fabricate, for example, carbon nanofoam.[18] Further examples of $sp^2$-rich magnetic carbon materials include functionalized carbon nanotubes,[19] metallo-fullerenes,[20] highly-oriented pyrolytic graphite,[21] and nano-graphite.[14] Additional studies have been carried out on iron, fluorine and boron implanted graphite[22] and graphene,[23] but have provided inconsistent results.[24] The extent of magnetization and the type of magnetism differ significantly in all these cases. Most of these magnetic carbons can be generally classified as either (i) carbon with artificially created dangling bonds, or (ii) carbon with a magnetic impurity.

The fabrication of pyrolytic magnetic carbon (PMC) reported here is a different and considerably more robust approach, because instead of inducing dangling bonds or adding any magnetic entity, we tune the microstructure of the carbon itself during its pyrolysis such that there is a build-up of unpaired electrons. Pyrolysis is a MEMS compatible process (used for carbon-MEMS fabrication[25]) that encompasses heat-treatment of carbon-rich polymer precursors in an inert environment. The standard pyrolysis is carried out at temperatures > 800 °C, which yields an amorphous glass-like carbon.[26] In this polymer-to-carbon conversion process, 550–700 °C is the temperature range where the C–H bonds are broken and the pyrolyzing matrix is rendered highly disordered.[25] In this transition region the physical, chemical and electronic properties of the carbonizing material undergo a sudden and dramatic change due to the formation of a high fraction of unsaturated bonds. For PMC fabrication, we performed pyrolysis at 620 °C to freeze this disordered state of the carbon matrix, thus resulting in a material with intrinsic dangling bonds. Magnetism in polymer derived carbons due to unpaired electrons was also noted in the early 1960s,[26] however, no follow-up in terms of applicability of such materials could be traced. Recently, advanced carbonizable photoresists such as epoxy resins have opened up numerous microfabrication possibilities,[25] and here we have characterized the most popular MEMS compatible polymers for their magnetic properties after their conversion into PMC.

Three polymers, SU-8,[25] mr-6000-NIL,[27] and polyacrylonitrile (PAN)[25] were selected to demonstrate that (i) the magnetism in PMC is independent of the chemical composition of the starting material, and is not resulting from any impurity (such as photoinitiator salts, H or N), (ii) the precursors can be patterned into a variety of shapes by photolithography or electrospinning prior to their conversion into PMC. We tested the PMC manufacturing method for thin-films, isolated and bulk micro/ nanofibers, and powder samples. The magnetic properties of PMC, probed mainly by electron paramagnetic resonance (EPR) spectroscopy, were highly consistent for all types of samples.

First, we determined the specific pyrolysis temperature and dwell time (the duration for which a sample is maintained at its pyrolysis temperature) that yield PMC with the strongest paramagnetic resonance from epoxy-based photoresists. For this purpose, EPR was performed on PMC thin-films pyrolyzed in a temperature series spanning the entire transition region (550–700 °C). The EPR signal intensities revealed that the highest unpaired electron-spin concentration was produced at a pyrolysis temperature of 600–620 °C (Figure 1 a). Subsequently, the samples pyrolyzed at 620 °C for dwell times ranging between 6 minutes to 5 hours were analyzed by EPR. The results, presented in Figure 1 b, indicate that a dwell time between 30 to 60 minutes yields PMC with the highest spin concentration. In the work reported here all samples were pyrolyzed at 620 °C for 1 hour to obtain a high spin concentration while still allowing sufficient time for the release of any volatile impurities.

EPR spectra were obtained for a selection of electrospun or photopatterned PMC samples. PMC powder, derived from SU-8, exhibited an intense isotropic signal with a peak-to-peak linewidth of ~1 mT, indicating a high radical concentration devoid of any noticeable anisotropic interactions (Figure 1 c). PMC nanofibers derived from electrospun PAN produced a strong EPR signal with a 0.15 mT linewidth, significantly narrower than the bulk SU-8 sample (Figure 1 d). Nanofibers have a rather well defined morphology and a lower packing density compared to powder samples, resulting in a lower contribution to the EPR linewidth coming from electronic dipolar interactions.[28]

PMC may contain 1–2% H impurity,[25] which raises the question whether the EPR signal results from electrons residing on H atoms. In that case one should observe a characteristic hyperfine interaction due to electron spin coupling to the nuclear spin of H. To probe the presence of such hyperfine coupling, the mass of the PMC samples was reduced, which would concomitantly reduce the large peak linewidth typical of powder samples and reveal the hyperfine couplings, if any. The acquired EPR spectra from these samples displayed the expected decrease in signal-to-noise ratio, but retained their lineshapes (Figures 1 e, f). The absence of any *g*-tensor anisotropies or hyperfine couplings indicates that the EPR signal originates from carbon-associated electrons, and not from any impurities.

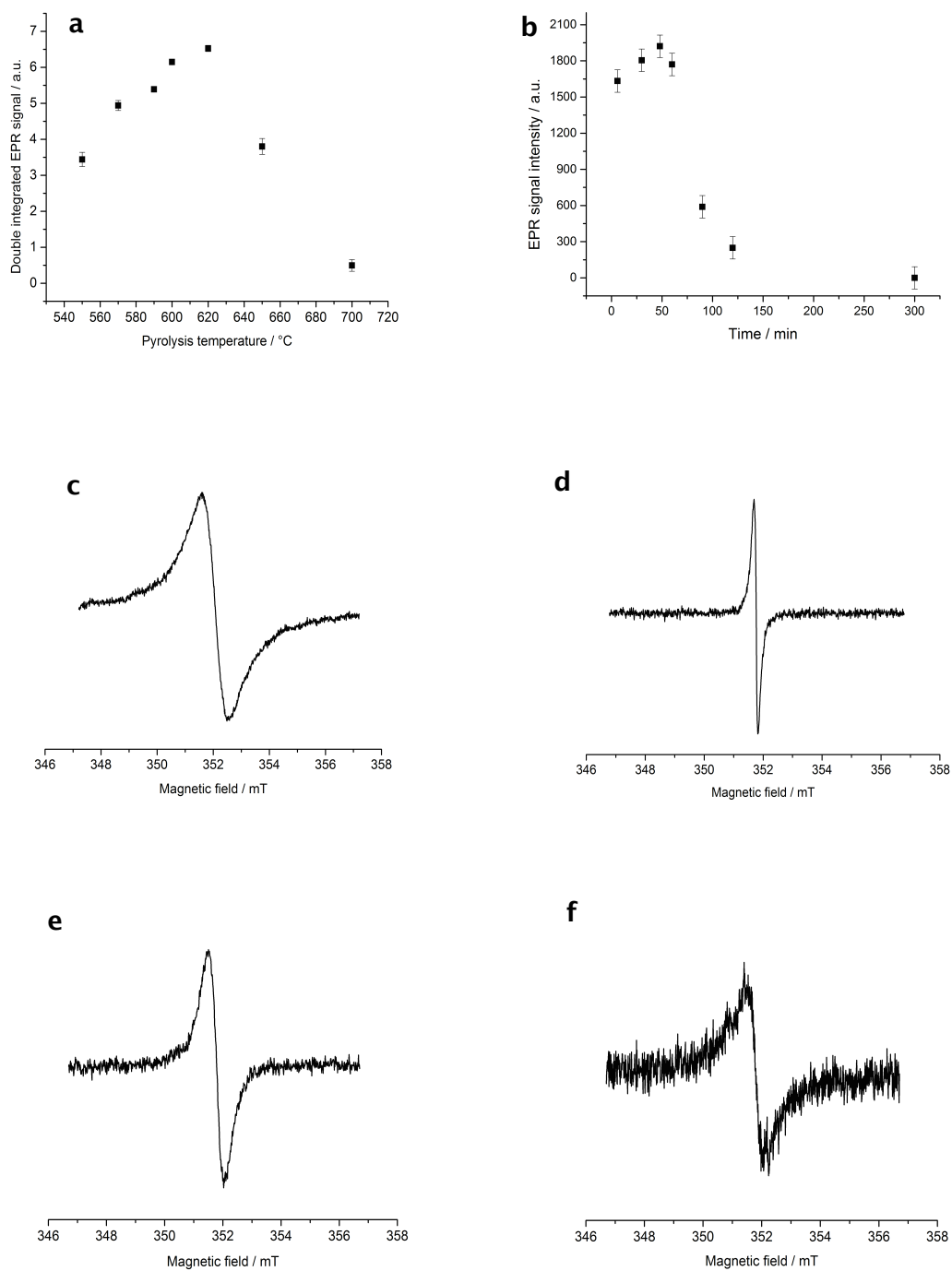

Figure 1. (a) EPR signal (second integral) of PMCs obtained at different pyrolysis temperatures ranging between 550-700 °C. (b) EPR signal of PMCs obtained at 620° C at different pyrolysis dwell times. (c) EPR spectrum of PMC powder sample, (d) EPR spectrum of PMC nanofibers (bulk), (e) EPR spectrum of PMC thin-films, (f) EPR spectrum of isolated PMC fibers. All EPR spectra were measured with an X-band spectrometer using a super-high-Q resonator. The microwave power was set to 0.6 mW, the conversion time to 163 ms, the number of scans to 3 and the modulation amplitude to 0.1 mT.

The isotropic *g*-values for various pyrolysis temperatures obtained from different starting materials employing different micro- and nanofabrication methodologies are compiled in Table 1. A characteristic free electron *g*-value of 2.0002 was obtained for all samples (within the experimental error). This suggests that the pyrolysis temperature only changes the radical concentration, but has no influence on the chemical environment of the radicals generated during the process. The highly consistent *g*-values observed here further substantiate the fact that the EPR signals originate from the same chemical entity in all samples, *i.e.*, from carbon-associated unpaired electron spins.

**Table 1: *g*-factors for various magnetic carbon samples**

| Polymer | Sample type | Pyrolysis Temperature ($^{o}$C) | *g*- factor |
|---|---|---|---|
| SU-8 | Powder | 620 | $2.0009 \pm 0.0006$ |
|  | Isolated nanofibers on Si | 620 | $2.0002 \pm 0.0006$ |
| PAN | Bulk nanofibers | 620 | $2.0002 \pm 0.0006$ |
|  | Isolated nanofibers on Si | 620 | $2.0002 \pm 0.0006$ |
| mr-NIL-6000 | Thin-film on Si | 550 | $2.0002 \pm 0.0006$ |
|  |  | 570 | $2.0000 \pm 0.0006$ |
|  |  | 590 | $1.9999 \pm 0.0006$ |
|  |  | 600 | $1.9996 \pm 0.0006$ |
|  |  | 620 | $1.9998 \pm 0.0006$ |
|  |  | 650 | $2.0002 \pm 0.0006$ |
|  |  | 700 | $2.0002 \pm 0.0006$ |
|  |  | 900 | $2.0007 \pm 0.0006$ |

Typical fitted first integrals of the EPR spectra of PMC exhibit a near-perfect Lorentzian lineshape with negligible Gaussian contributions (Figure 2 a). This again confirms that the material has a high degree of homogeneity in terms of the distribution of the spins, and rules out the existence of any hyperfine couplings and/ or *g*-anisotropies.

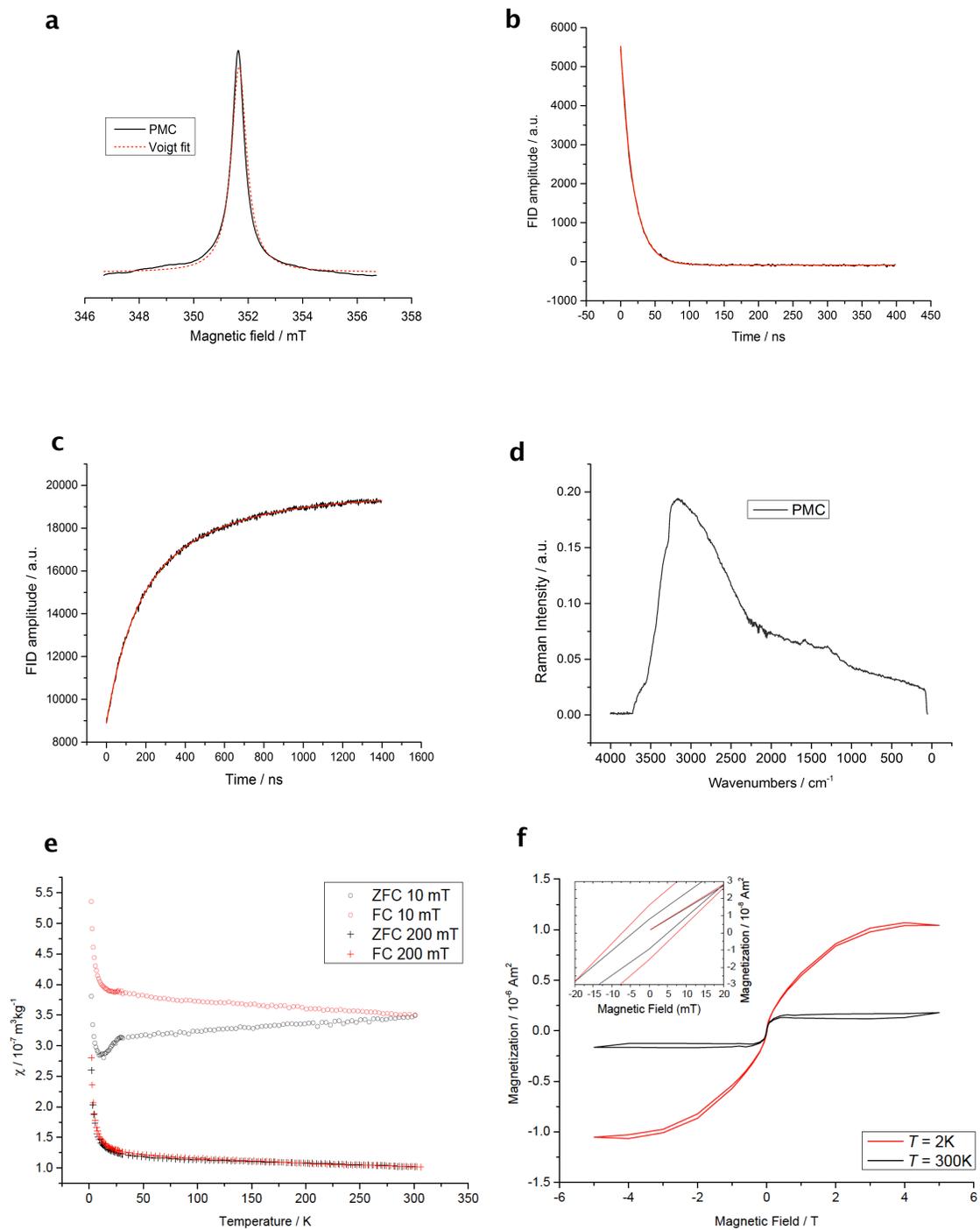

**Figure 2. Magnetic and microstructural characterization of PMC (a)** Typical fitted first integral of EPR signal for PMC. **(b, c)** Free Induction Decay of PMC for $T_2$ and $T_1$ calculations respectively, **(d)** Raman spectrum of PMC. **(e, f)** Magnetic susceptibility curves for PMC: **(e)** Field-cooled and zero-field-cooled magnetic carbon samples at 0.01 and 0.2 T, **(f)** magnetization curves obtained at 2 K and 300 K. Inset: a close-up plot of the hysteresis loops near the plot origin. All diamagnetic contributions from the sample holder and any diamagnetism from the sample itself have been subtracted from the data points of plot (f).

The electron spin-lattice and spin-spin relaxation times ($T_1$ and $T_2$, respectively) of PMC at room temperature were determined by analyzing the peak-to-peak EPR linewidths as a function of the microwave power[28] (Table 2). Evidently, the relaxation times for PMC are quite short. To obtain $T_2$ at lower temperatures, a free induction decay (FID) after a π/2 pulse for powder samples was acquired at 80 K. The temperature dependence of $T_1$, was evaluated by inversion recovery experiment.[28] The FID for $T_2$ and $T_1$ were fitted using a mono- and double-exponential decay respectively (Figures 2 b, c). As expected, at 80 K, $T_2$ remains mostly unchanged (23 ns), while $T_1$ increases to ~270 ns. These observations are indicative of strong spin-spin exchange interactions within the material, which are possible because of the high radical concentration [(3.97 ± 0.8) × $10^{17}$ spins/ mg].

Table 2: Electron $T_1$ and $T_2$ relaxation times for PMC obtained at room-temperature from various PMC samples

| Sample type | Unsaturated peak-to-peak linewidth (mT) | $T_1$ (ns) | $T_2$ (ns) |
|---|---|---|---|
| Thin-film (mr-6000-NIL) | 0.311 ± 0.001 | 17.87 ± 0.04 | 21.13 ± 0.05 |
| Nanofibers bulk (PAN) | 0.450 ± 0.001 | 25.87 ± 0.03 | 14.60 ± 0.02 |
| Powder (SU-8) | 1.120 ± 0.001 | 19.53 ± 0.02 | 58.46 ± 0.05 |

To investigate the microstructure of PMC, a Raman spectrum for the powder sample was obtained. As it can be observed in Figure 2 d, the spectrum is broad and nearly featureless in the range from 0 to ~3700 cm$^{-1}$ wavenumbers, devoid of any characteristic C–C stretching peaks for disordered or graphitic carbon moieties (*D* and *G* bands at ~1350 and 1680 cm$^{-1}$ respectively). Other plausible Raman peaks in disordered carbons[29] were also not detectable. The contributions from $sp^2$ carbon atoms in the spectrum are comparatively small and thus the material is expected to be mainly composed of *sp* hybridized, or possibly unhybridized (ground state) carbon atoms. While a long excitation wavelength ($\lambda$ = 1064 nm) was used to excite the sample during the Raman spectroscopic study to avoid any fluorescence in the visible region, one cannot rule out the possibility of absorption of the infrared wavelengths by the material. If Raman scattering is induced close to such electronic transitions, the associated emissions may mask parts of the Raman spectrum.

PMC was also subject to magnetic susceptibility measurements to test its characteristic magnetic response at various applied magnetic fields and temperatures. Magnetic susceptibilities extracted from the zero-field-cooled (ZFC) and field-cooled (FC) magnetization measurements are plotted as a function of the temperature in Figure 2 e. At $B$ = 0.01 T, one observes a clear splitting between the ZFC and FC curves, while at $B$ = 0.2 T the two curves are virtually identical. At both field strengths a steep monotonic decrease of $\chi$ on increasing the temperature from 4 to 10 K was observed, which is attributed to the fraction of paramagnetic centers in PMC, *i.e.*, the component originating from the localized electrons on carbon radicals.

The subsequent increase in the magnetization above $T$ = 10 K in the 0.01-T ZFC curve and the intersection of ZFC and FC curves at 300 K can be rationalized assuming the presence of magnetically correlated electron spins within nano-sized regions. This fraction of interacting spins in bulk PMC may exhibit either a superparamagnetic behavior with a blocking temperature above 300 K and a broad distribution of anisotropy energy barriers, or a partly-frustrated magnetic order with a glass temperature close to room temperature. The sudden increase in the ZFC magnetization curve between 10 K and 25 K is indicative of a currently unknown additional phenomenon.

The hysteresis loops at 300 K and 2 K (Figure 2 f) support the hypothesis of a not entirely paramagnetic phase: the curve at 300 K shows saturation at an applied magnetic field of approximately 0.5 T, whereas at 2 K the magnetization reaches saturation at 4 T. This is consistent with the paramagnetic component being dominant at low temperatures, while at high temperatures the behavior is closer to that of a ferromagnet or a superparamagnet. Values for the coercive field are $B_C = 4.5$ mT and $B_C = 7.5$ mT at $T = 300$ K and $T = 200$ K, respectively (see inset of Figure 2 f). The finite but low values found for $B_C$ are in line with values typically recorded for assemblies of superparamagnetic clusters.[30] Further investigations to better understand this superparamagnetic-like behavior in PMC are ongoing.

EPR measurements were also performed on a powder sample kept under 5 mbar pressure at 120 °C for 4 hours to ensure that the signal does not result from adsorbed molecular oxygen. The EPR signals for the same sample before and after degassing were almost identical. In addition, this experiment revealed the stability of PMC's magnetic properties during extended high-temperature exposure, a quality that is highly desirable for MEMS fabrication, where multiple processing steps at elevated temperatures are inevitable.

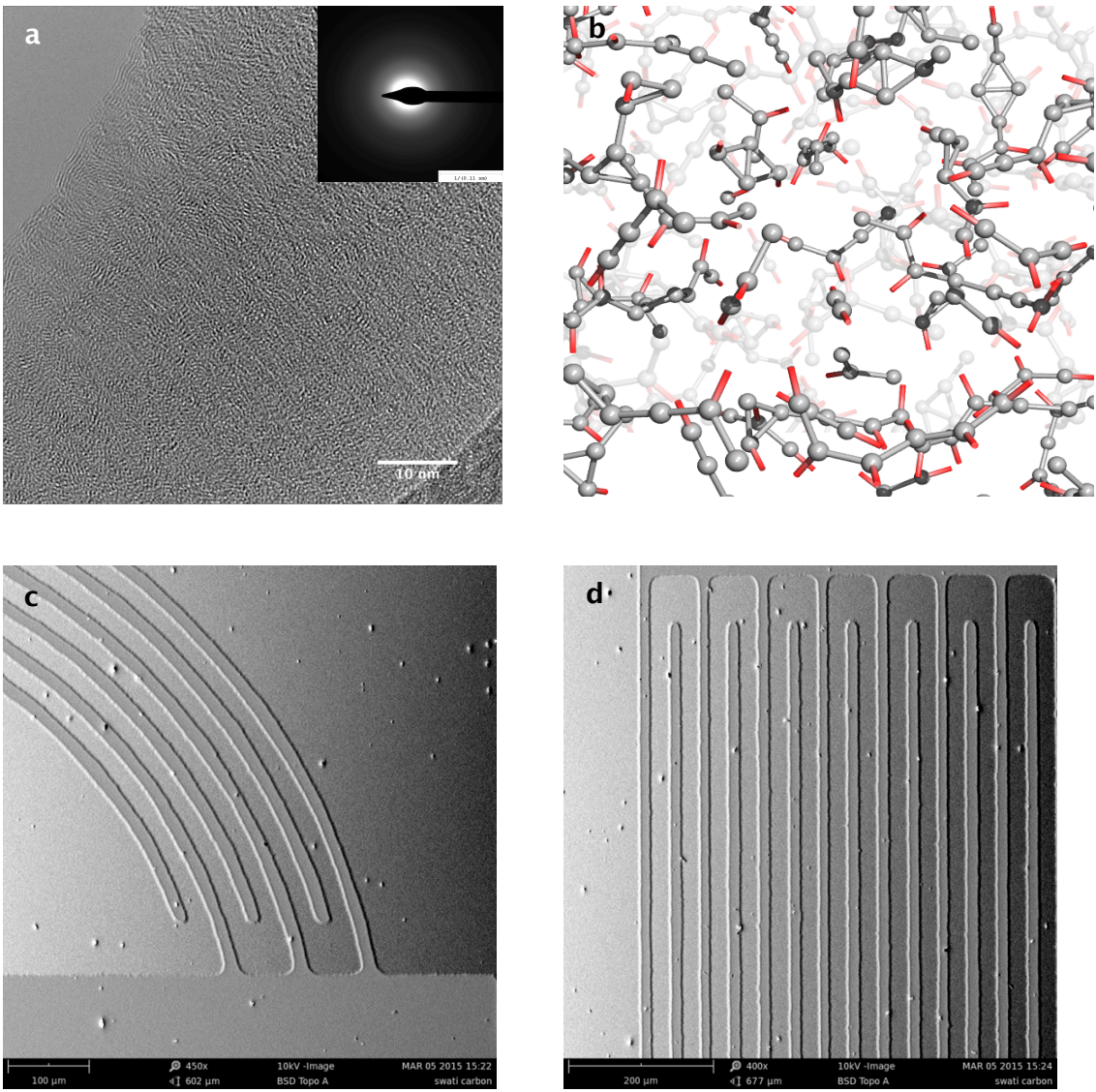

**Figure 3. Electron microscope images and pictorial model of PMC (a) HR-TEM image (inset: electron diffraction pattern) of SU-8 derived PMC powder. (b) Proposed model of the atomic arrangement in magnetic carbon. The dangling bonds are represented in red color. (c, d) SEM micrographs of test patterns fabricated in PMC.**

The high-resolution transmission electron microscope (HR-TEM) investigations of PMC revealed a primarily amorphous structure with a minor ordered component at the edge (Figure 3 a). The diffraction pattern (inset of Figure 3 a) suggests that the material is highly amorphous and does not feature any long-range order. Based on the Raman and the HR-TEM data, we propose a pictorial model for the atomic arrangements in PMC as illustrated in Figure 3 b. Finally, photopatterned SU-8 test structures were converted into PMC in order to demonstrate its MEMS fabrication capabilities (Scanning Electron Microscope images in Figures 3 c, d).

In summary, we have developed a novel, paramagnetic form of carbon through a facile wafer-scale process via the low-temperature pyrolysis of lithographically patterned precursor polymers. The fabrication techniques for PMC are not limited to photolithography and electrospinning; one can also pattern the precursor polymers using nano-imprint lithography, two-photon lithography, soft-lithography *etc.*, on a variety of substrates. PMC can be of a great value to researchers investigating the mechanism of carbon magnetism, as well as to those who envision fascinating engineering applications with magnetic carbon.

**Experimental methods**

*Fabrication*

SU-8 (MicroChem, MA, USA) or mr-6000-NIL (micro resist technology GmbH, Germany) was photopatterned onto a Si substrate. Alternatively, a 10% solution (by weight in dimethylformamide) of PAN (Sigma Aldrich) was electrospun to obtain the desired nanofibers. These samples were heated in a tube furnace (Carbolite Gero GmbH, Germany) under 0.8 L/min flow of $N_2$ to various reported pyrolysis temperatures at a ramp rate of 5 °C/min. The samples were maintained at their respective pyrolysis temperatures for 1 hour and were naturally cooled to room temperature before further analyses.

*EPR*

EPR measurements were carried out on a Bruker EMX X-Band spectrometer at a microwave frequency of ~9.4 GHz and a magnetic field around 350 mT, with a 4122-SHQ-E BRUKER microwave resonator. For temperature control a Bruker ER4131VT digital temperature control system with controllable temperature range from 100 K to 500 K was used. The microwave power was set to 0.6 mW and the magnetic-field modulation amplitude was 0.1 mT. For the relaxation time measurements EPR spectra were recorded at different microwave attenuations ranging from 15 to 51 dB (corresponding to microwave powers of 6315 to 1.586 µW) in 3-dB-steps for all samples. The resonator quality factor was 5400 in all measurements. The pulsed EPR measurements were carried out on a ELEXSYS E580 spectrometer equipped with a Bruker X-MD4 resonator. For determining the spin concentration, numerical integration of the PMC EPR signal was compared with that of a $CuSO_4$ standard recorded at the same temperature.

*Fit functions*

The fit function used in Figure 2 a (Lorentzian and Gaussian fraction quantification of the first integral of the EPR signal) was a Voigt function. For the relaxation times (Figures 2 b, c), we used mono- and double-exponential decay functions.

*Microstructural characterization and imaging*

HR-TEM images and diffraction patterns for the suspension of PMC powder in deionized water were obtained on a Philips CM200 FEG/ST TEM system. A Bruker VERTEX 70 Raman spectrometer equipped with FTIR was used for analyzing a powdered PMC sample. SEM images of the photopatterned PMC structures were acquired on a Phenom ProX Desktop SEM system.

*SQUID Magnetometer*

Static magnetic measurements were performed on a MPMS-XL7 Quantum Design superconducting quantum interference device (SQUID) at applied static magnetic fields $B = 10$ mT and $B = 200$ mT, in the temperature range 2-300 K. The hysteresis loops were recorded within the field range $-5 < B < 5$ T at temperatures $T = 2$ K and $T = 300$ K. The diamagnetic contribution from the sample holder was subtracted from all magnetization measurements.

**Acknowledgements**

The authors (S.S., N.M., L.B. & J.G.K.) gratefully acknowledge the European Research Council (ERC) for funding via the advanced grant no. 290586 NMCEL. In addition, they convey their sincere thanks Mr. Tarek Al Said (University of Freiburg, Germany) for contributions to EPR analyses, Prof. Alessandro Lascialfari (University of Pavia, Italy) for magnetic susceptibility measurements, Mr. Thomas Vent-Schmidt (University of Freiburg, Germany) for Raman spectroscopy, and Prof. Dr. Thomas Hanemann (University of Freiburg, Germany) for providing access to the furnace used for this work.

**Author contributions**

**S.S.** performed first proof of principle, material development and process optimization, sample fabrication, microstructural characterization and imaging, data analysis, and experiment planning. **A.M.R.** performed EPR experiment planning, and conducted EPR

# References

1. Rode, A. *et al.* Unconventional magnetism in all-carbon nanofoam. *Phys. Rev. B* **70**, (2004).

2. Bandow, S. Magnetic properties of nested carbon nanostructures studied by electron spin resonance and magnetic susceptibility measurements. *J. Appl. Phys.* **80**, 1020–1027 (1996).

3. Ramirez, A. P. *et al.* Magnetic Susceptibility of Molecular Carbon: Nanotubes and Fullerite. *Science* **265**, 84–86 (1994).

4. Parkansky, N. *et al.* Magnetic properties of carbon nano-particles produced by a pulsed arc submerged in ethanol. *Carbon* **46**, 215–219 (2008).

5. Esquinazi, P. *et al.* Induced Magnetic Ordering by Proton Irradiation in Graphite. *Phys. Rev. Lett.* **91**, (2003).

6. Ohldag, H. *et al.* π-Electron Ferromagnetism in Metal-Free Carbon Probed by Soft X-Ray Dichroism. *Phys. Rev. Lett.* **98**, (2007).

7. Gibbs, M. R. J., Hill, E. W. & Wright, P. J. Magnetic materials for MEMS applications. *J. Phys. Appl. Phys.* **37**, R237–R244 (2004).

8. Niarchos, D. Magnetic MEMS: key issues and some applications. *Sens. Actuators Phys.* **106**, 255–262 (2003).

9. Makarova, T. L. *et al.* Retraction: Magnetic carbon. *Nature* **440**, 707–707 (2006).

10. Ohldag, H. *et al.* The role of hydrogen in room-temperature ferromagnetism at graphite surfaces. *New J. Phys.* **12**, 123012 (2010).

11. Červenka, J., Katsnelson, M. I. & Flipse, C. F. J. Room-temperature ferromagnetism in graphite driven by two-dimensional networks of point defects. *Nat. Phys.* **5**, 840–844 (2009).

12.     Wagoner, G. Spin Resonance of Charge Carriers in Graphite. *Phys. Rev.* **118**, 647–653 (1960).

13.     Rao, S. S., Stesmans, A., Kosynkin, D. V., Higginbotham, A. & Tour, J. M. Paramagnetic centers in graphene nanoribbons prepared from longitudinal unzipping of carbon nanotubes. *New J. Phys.* **13**, 113004 (2011).

14.     Wakabayashi, K. & Harigaya, K. Magnetic Structure of Nano-Graphite Möbius Ribbon. *J. Phys. Soc. Jpn.* **72**, 998–1001 (2003).

15.     Li, S., Ji, G. & Lü, L. Magnetic Carbon Nanofoams. *J. Nanosci. Nanotechnol.* **9**, 1133–1136 (2009).

16.     Höhne, R. *et al.* Magnetism of pure, disordered carbon films prepared by pulsed laser deposition. *J. Magn. Magn. Mater.* 272-276, E839–E840 (2004).

17.     Sepioni, M. *et al.* Limits on Intrinsic Magnetism in Graphene. *Phys. Rev. Lett.* **105**, (2010).

18.     Rode, A. V. *et al.* Strong paramagnetism and possible ferromagnetism in pure carbon nanofoam produced by laser ablation. *J. Magn. Magn. Mater.* **290-291**, 298–301 (2005).

19.     Wang, H. *et al.* Fe nanoparticle-functionalized multi-walled carbon nanotubes: one-pot synthesis and their applications in magnetic removal of heavy metal ions. *J. Mater. Chem.* **22**, 9230 (2012).

20.     Adiseshaiah, P. *et al.* A Novel Gadolinium-Based Trimetasphere Metallofullerene for Application as a Magnetic Resonance Imaging Contrast Agent: *Invest. Radiol.* **48**, 745–754 (2013).

21.     Dubman, M. *et al.* Low-energy and SQUID evidence of magnetism in highly oriented pyrolytic graphite. *J. Magn. Magn. Mater.* **322**, 1228–1231 (2010).

22.     Höhne, R. *et al.* The influence of iron, fluorine and boron implantation on the magnetic properties of graphite. *J. Magn. Magn. Mater.* **320**, 966–977 (2008).

23. Kim, H.-J. & Cho, J.-H. Fluorine-induced local magnetic moment in graphene: A hybrid DFT study. *Phys. Rev. B* **87**, (2013).

24. Nair, R. R. *et al.* Spin-half paramagnetism in graphene induced by point defects. *Nat. Phys.* **8**, 199–202 (2012).

25. Sharma, S. Microstructural tuning of glassy carbon for electrical and electrochemical sensor applications. (Ph.D. Thesis, University of California, Irvine, CA, USA, 2013).

26. Jenkins, G. M. *Polymeric carbons--carbon fibre, glass and char*. (Cambridge University Press, 1976).

27. Schuster, C., Reuther, F., Kolander, A. & Gruetzner, G. mr-NIL 6000LT – Epoxy-based curing resist for combined thermal and UV nanoimprint lithography below 50°C. *Microelectron. Eng.* **86**, 722–725 (2009).

28. Poole, C. P. *Electron spin resonance: a comprehensive treatise on experimental techniques*. (Wiley, 1983).

29. Schwan, J., Ulrich, S., Batori, V., Ehrhardt, H. & Silva, S. R. P. Raman spectroscopy on amorphous carbon films. *J. Appl. Phys.* **80**, 440 (1996).

30. Kolhatkar, A., Jamison, A., Litvinov, D., Willson, R. & Lee, T. Tuning the Magnetic Properties of Nanoparticles. *Int. J. Mol. Sci.* **14**, 15977–16009 (2013).